\documentclass[11pt]{bull-l}
\usepackage{amssymb}

\begin{document}
\address{Jan Philip Solovej\\ University of Copenhagen}
\email{solovej@math.ku.dk}

\noindent {\it Stability of Matter in Quantum Mechanics}, by Elliott
H.\ Lieb and Robert Seiringer, Cambridge University Press, Cambridge,
2010, xv+293 pp, ISBN 978-0-521-19118-0.
\let\thefootnote\relax \bigskip

Stability of matter is a fundamental fact about the nature of ordinary
matter. In essence it says that macroscopic objects exist! It is at
the same time a rigorous mathematical statement in the theory of
quantum mechanics. I will describe its precise meaning below. It is
somewhat surprising that stability of matter is not a subject treated
in standard physics textbooks. It is however one of the most
celebrated results and a cornerstone in mathematical physics.  The
book under review {\it Stability of Matter in Quantum Mechanics} by
Lieb and Seiringer is the first to give a complete and thorough
account of stability of matter.  I will begin with an overview of the
subject itself.

The reason stability of matter is not treated in physics textbooks is
not because of its lack of importance, in fact, what could be more
important? More likely, the reason is that it is not easy to derive.
In contrast to most other results in mathematical physics there was,
to the best of my knowledge, no heuristic derivation of stability of
matter prior to the rigorous proof of the theorem, which appeared in
1967 in the seminal work of Dyson and Lenard~\cite{DL}. Even Onsager's
paper~\cite{O} which is probably the very first to address the issue
was mathematically correct and presented ideas used in many later
works.

Stability is an important concept in physics and the notion is used in
many contexts. One of the triumphs of the theory of quantum mechanics
is that it explains the stability of atoms. The puzzling question
settled by quantum mechanics is why the electrons in the atom do not
simply collapse on top of the atomic nucleus due to their mutual
electrical attraction. There are two ways to formulate this
problem. We might ask why there is dynamic stability, i.e., why the
motion is well-defined for all times independently of the initial
condition. Or we might alternatively ask why there is energetic
stability, i.e., why the total energy cannot be arbitrarily negative,
which it would if the electrons were arbitrarily close to the
nucleus. This is indeed what may happen in classical mechanics, where
we have neither energetic nor dynamic stability.

It turns out that in quantum mechanics energetic stability implies
dynamic stability. In technical terms if the energy is bounded below
there is a natural realization (the Friedrichs extension) of the
energy operator, as a self-adjoint operator i.e., the Hamiltonian, on
a Hilbert space. This operator generates the dynamics.  Hence a study
of stability in quantum mechanics may focus on energetic stability.
This is the topic of the book by Lieb and Seiringer.

The energetic stability of atoms, or more precisely of the hydrogen atom
(an atom with one electron), is usually explained, at least heuristically,
in the first few pages of most textbooks in quantum mechanics.  The
explanation is based on {\it the uncertainty principle}. Likewise, 
stability of the hydrogen atom is proved in the introductory chapter
of the book by Lieb and Seiringer. It is pointed out here, however, that,
contrary to what is stated in most physics texts, the famous
Heisenberg formulation of the uncertainty principle is, in fact, not
very useful in order to conclude stability. For this purpose the {\it
  Sobolev inequality} is a better formulation of the uncertainty
principle and is used in the book by Lieb and Seiringer to prove 
stability of the hydrogen atom.

Energetic stability, i.e., the fact that there is a lower bound to the
energy, is referred to in the book as {\it stability of the first
  kind}.  Stability of matter also called {\it stability of the second
  kind} is a more complicated notion relating to the energy of
macroscopic systems.  Individual atoms or molecules are relatively
small systems with a few degrees of freedom. Macroscopic matter,
however consists of an enormous amount of atoms, i.e., it is made out
of a macroscopic number of nuclei and electrons. As an example one
gram of hydrogen consists of approximately $6\cdot10^{23}$ (Avogadros'
number) hydrogen atoms. Stability of the first kind only states that
the energy of such a system is not arbitrarily negative. It does not
address the issue of how negative it may be depending on the size of
the system, e.g., measured by the number of particles. For macroscopic
systems, however, it is important that the dependence of the energy on
the size of the system is at most linear.  The energy of twice an
amount of a substance should be essentially twice the energy of the
amount itself.  This is {\it stability of matter}.  It is closely
related to the {\it extensivity of matter}, i..e, that the volume of a
substance grows proportional to its quantity, otherwise a macroscopic
number of particles would not take up a macroscopic volume. As obvious
as this may sound it is difficult to prove.
 
Contrary to stability of the first kind stability of the second kind
does not follow from the uncertainty principle alone. It requires also
the {\it Pauli-exclusion principle}, i.e., the fact, to be explained
below, that electrons are fermions and thus cannot occupy the same
one-particle states.  Without the exclusion principle stability of
matter fails. In fact, as first noted by Dyson~\cite{D}, the energy of
such a system would have a super-linear behavior as a function of
particle number and the volume would, indeed, {\it decrease}; {\it
  more} particles would take up {\it less} space.

It was mentioned above that stability of matter is usually not treated
in physics textbooks.  There is however another case of stability due
to the Pauli-exclusion principle which is known to any physicist. This
is Chandrasekhar's famous theory~\cite{C} (for which he got the Nobel
prize in 1983)  of gravitational stability and instability of stars in
their late evolutionary state as white dwarfs. Chandrasekhar's theory
was given a rigorous formulation in \cite{LTc,LY} and this is also
covered in Lieb and Seiringer's book.

Besides being a problem of basic physical importance, the study of
stability of matter leads to a wealth of beautiful mathematics.
Topics such as variational calculus, potential theory, operator
theory, spectral theory, Sobolev inequalities and phase space analysis
need to be brought together in order to arrive at a proof of stability
of matter.

Let me briefly review the precise formulation of stability of matter
and as a guide to the reader of the book indicate the main steps in
its proof.

Matter is described as consisting of electrons and nuclei. All the
electrons are identical with the same mass $m$ and negative charge
$-e$. The nuclei may be different and have different masses and
(positive) charges. The charge of a nucleus is $Ze$ where the integer
$Z$ is the atomic number of the nucleus. The smallest nucleus is the
hydrogen nucleus (a single proton) with $Z=1$ and all naturally
existing nuclei have $Z\leq92$ corresponding to the elements in the
periodic table.

Imagine that we have $N$ electrons and $M$ nuclei with atomic numbers
$\underline{Z}=(Z_1,\ldots,Z_M)$.  Let $E_{N,M}(\underline{Z})$ be the
smallest possible (actually the infimum) energy of such a system.  It
depends on the nuclear charges, their masses, the mass and charge of
the electron, and Planck's constant $\hbar$ (this is really Planck's
constant divided by $2\pi$). Stability of the first kind is the claim
that $E_{N,M}(\underline{Z})$ is finite (not negative
infinity). Stability of matter states that
\begin{equation}\label{eq:stabmat}
  E_{N,M}(\underline{Z})\geq -\Xi(Z)(N+M),
\end{equation}
where the constant $\Xi(Z)$ depends only on
$Z=\max\{Z_1,\ldots,Z_M\}$, Planck's constant, and the mass and charge
of the electron, but {\it not} on the masses of the
nuclei. Establishing stability of matter with a constant independent
of the masses of the nuclei is physically important. Nuclei are much
heavier than electrons and the energy per particle should not diverge
as the masses tend to infinity. In other words we might as well think
of the worst case scenario when the masses of the nuclei are all
infinite.  This is the case referred to as static nuclei.

To give the precise definition of $E_{N,M}(\underline{Z})$ we
introduce the 3-dimensional coordinates of the electron positions
$\underline{X}=({\bf x}_1,\ldots,{\bf x}_N)\in{\mathbb R}^{3N}$, and
the nuclear positions $\underline{R}=({\bf R}_1,\ldots,{\bf
  R}_M)\in{\mathbb R}^{3M}$. The state of the electrons is described by a
complex valued wave function $\psi(\underline{X},\underline{\sigma})
$, where $\underline{\sigma}=(\sigma_1,\ldots,\sigma_N)$ denote the
internal spin degrees of freedom. Each $\sigma_i$ can take $q$ values.
For physical electrons the spin is $1/2$ corresponding to $q=2$, but
in the discussion here $q$ could be any positive integer.  The wave
function should be normalized, i.e,
$\sum_{\underline{\sigma}}\int_{{\mathbb
    R}^{3N}}|\psi(\underline{X},\underline{\sigma})|^2d\underline{X}=1$,
where $d\underline{X}=d{\bf x}_1\cdots d{\bf x}_N$.

The important Pauli exclusion principle can now be formulated as the
requirement that the wavefunction is {\it fermionic}, which means that
it is antisymmetric under the interchange of $({\bf x}_i,\sigma_i)$
and $({\bf x}_j,\sigma_j)$ for any $i\ne j$.  

The energy consists of two parts a {\it kinetic energy}, which is
\begin{equation*}
T_\psi=\frac{\hbar^2}{2m}\sum_{i=1}^N\sum_{\underline{\sigma}}\int_{{\mathbb R}^{3N}}
|\nabla_{{\bf x}_i}\psi(\underline{X},\underline{\sigma})|^2d\underline{X}
\end{equation*}
and a {\it potential energy} which is 
\begin{equation*}
V_\psi(\underline{R})=\sum_{\underline{\sigma}}\int_{{\mathbb
    R}^{3N}}V_{\rm
  C}(\underline{X},\underline{R})|\psi(\underline{X},\underline{\sigma})|^2d\underline{X},
\end{equation*}
where we have introduced the electrostatic Coulomb potential
\begin{equation*}
  V_{\rm C}(\underline{X},\underline{R})=-\sum_{i=1}^N\sum_{j=1}^M\frac{Z}{|{\bf x}_i-{\bf R}_j|}
  +\sum_{1\leq i<j\leq N}\frac1{|{\bf x}_i-{\bf x}_j|}+\sum_{1\leq i<j\leq M}\frac{Z^2}{|{\bf R}_i-{\bf R}_j|}.
\end{equation*}
For simplicity we here consider the case where all nuclear charges are
equal, i.e, equal to the maximal value $Z$.  There is a monotonicity
argument showing that this is, indeed, the worst case.  Finiteness of
the kinetic energy $T_\psi$ implies that $\psi$ belongs to the Sobolev
space $H^1({\mathbb R}^{3N})$.  For such a function all terms in
$V_\psi$ are also finite.  The precise definition of the energy is
then
\begin{equation*}
E_{N,M}(\underline{Z})=\inf\left\{\ T_\psi +
  e^2V_\psi(\underline{R}):\ \underline{R}\in{\mathbb R}^{3M}, \psi\in
  H^1\hbox{ fermionic, normalized}\right\}.
\end{equation*} 
Note that the static
nuclei are described only through their positions $\underline{R}$,
which are optimized in order to minimize the energy.

Stability of matter (\ref{eq:stabmat}) can be derived from
inequalities on the kinetic energy $T_\psi$ and the Coulomb potential
$V_{\rm C}$.  The first fundamental inequality is the
Lieb-Thirring~\cite{LT1} kinetic energy estimate for normalized
fermionic wavefunctions $\psi$ with $q$ spin states
\begin{equation*}
  T_\psi\geq \frac{\hbar^2}{2m}\frac{K}{q^{2/3}}\int_{{\mathbb R}^3}\rho_\psi({\bf x})^{5/3}d{\bf x}.
\end{equation*}
Here we have introduced the electronic density
\begin{equation*}
  \rho_\psi({\bf x_1})=N\sum_{\underline{\sigma}}\int_{{\mathbb
      R}^{3(N-1)}}|\psi(\underline{X},\underline{\sigma})|^2d{\bf
    x}_2\cdots d{\bf x}_N.
\end{equation*}
Note that the normalization condition on $\psi$ implies that
$\int_{{\mathbb R}^3}\rho_\psi({\bf x})d{\bf x}=N$.  A celebrated
(still unsolved)
conjecture of Lieb and Thirring~\cite{LT2} is that the best constant $K$
in the above inequality is obtained from the semiclassical expression
\begin{equation*}
K_{\rm CL}=\frac{(2\pi)^{-3}\int_{\{{\bf p}\in{\mathbb R}^3\ :\ |{\bf
      p}|\leq 1\}}{\bf p}^2d{\bf p}}{ \left((2\pi)^{-3}\int_{\{{\bf
        p}\in{\mathbb R}^3\ :\ |{\bf p}|\leq 1\}}1d{\bf
      p}\right)^{5/3}}=\frac35(6\pi^2)^{2/3}.
\end{equation*}

The second ingredient in deriving stability of matter is to control
the Coulomb potential energy.  There are several approaches to this
part. The simplest, which however does not lead to the best known
constant in (\ref{eq:stabmat}), uses an estimate by Baxter~\cite{B} on
the Coulomb potential.  It states that for all
$\underline{X}\in{\mathbb R}^{3N}$ and $\underline{R}\in{\mathbb
  R}^{3M}$ we have
\begin{equation*}
  V_{\rm C}(\underline{X},\underline{R})\geq-(2Z+1)\sum_{i=1}^NW_{\underline{R}}({\bf x}_i),\qquad
  W_{\underline{R}}({\bf x})=\max_{j=1,\ldots,M}\{|{\bf x}-{\bf R}_j|^{-1}\}.
\end{equation*}
Baxter proved this with probabilistic methods, but it can be derived~\cite{LYr}
using potential theory and elaborating on the original ideas of Onsager~\cite{O}.
Lieb and Seiringer give several stronger versions of this type of
electrostatic inequality. For the discussion here this original
version will suffice. The importance of the inequality is that the
Coulomb potential which contains terms depending on pairs of electron
coordinates is estimated by a sum of terms containing only individual
electron coordinates.  This leads to an estimate on the energy that
can be expressed entirely from the electron density
\begin{equation*}
  T_\psi + e^2V_\psi(\underline{R})\geq
  \frac{\hbar^2}{2m}\frac{K}{q^{2/3}}\int \rho_\psi({\bf x})^{5/3}d{\bf
    x}-e^2(2Z+1)\int\rho_\psi({\bf x})W_{\underline{R}}({\bf x})d{\bf x}
\end{equation*}
from which it is an easy exercise to derive (\ref{eq:stabmat}).

It turns out that (\ref{eq:stabmat}) holds also with $N+M$ replaced by
$M$, i.e., only the number of nuclei. The reason is that $N$ much
larger than $MZ$ would mean that the system is very far from being
electrically neutral, in fact, it would be very negatively charged and
this is not energetically favorable.  Such an argument sounds
intuitively simple but is, in fact, rather subtle and has been a very
active research area in mathematical physics and is still not fully
understood. It is often referred to as the {\it ionization problem}
because it may be rephrased as the question: what is the maximal
negative ionization of a system? Because of its implications to
stability of matter, Lieb and Seiringer use the opportunity to review
what is known about this intriguing problem and in particular prove
the stronger version of (\ref{eq:stabmat}).
 
We have briefly reviewed some of the basic ideas presented in great
details and with beautiful clarity in essentially the first half of
{\it Stability of Matter in Quantum Mechanics}.

On a historical note the book does not contain the original proof of
stability of matter by Dyson and Lenard~\cite{DL}. It is closer in
spirit to the later and more elegant approach of Lieb and
Thirring~\cite{LT1}. This latter derivation was, however, based on
Thomas-Fermi theory which the book chooses to circumvent.

The stability of matter discussed up to this point is for
non-relativistic quantum mechanics. Relativistic effects and in
particular the interaction with the electromagnetic field are
important phenomena.  The emission and absorption of light are
processes of basic importance to the structure of atoms and it has
been ignored in the discussion so far. Unfortunately, there is no
complete mathematical theory describing relativistic quantum mechanics
and the interaction of light and matter. Results on stability are
known in several approximate models and these are also described in
detail in the book. Although these models do not claim to be complete
they contain the basic feature, believed to be correct for all
relativistic models, that {\it instability} occurs in certain ranges
of the physical parameters. Extensions of stability of matter from the
non-relativistic setting is still a very active research area.

The last chapter in the book contains a proof of {\it existence of the
  thermodynamic limit}.  This refers to the fundamental property that
the energy, or for positive temperature systems {\it the free energy},
per volume is not only bounded but has a limit as the system size
tends to infinity. The first proof of this was due to Lieb and
Lebowitz~\cite{LL} and the proof in the book follows this original
approach.

Stability of matter may be considered a step towards the more
fundamental existence of the thermodynamic limit. Historically this
was how stability of matter was viewed~\cite{FR}, but over the decades
it has grown to be a subject in its own.

The subject is very much alive. Of particular interest to readers of
the book is a recent fairly elementary proof of the Lieb-Thirring
inequality which appeared \cite{R1,R2} after the publication of the
book.

Over the years there have been short reviews on stability of matter,
e.g.~\cite{Lr}, and the subject has been treated briefly in
mathematical physics texts such as \cite{S,T}.  A comprehensive
textbook on the subject useful to researcher and students alike is
long overdue.  The book {\it Stability of Matter in Quantum Mechanics}
is just that. A book that an experienced researcher in mathematics or
physics can use to learn the subject, a book that the expert in the
field must have, and a book that is well suited for a semester course
for graduate students. In particular, the book can serve well as an
introduction for mathematicians to quantum mechanics.

The book by Lieb and Seiringer presents physical ideas and concepts
with mathematical rigor. It is not a book only about mathematics nor a
book only about physics. It is a book about both. A book in
mathematical physics. 

Stability of matter is an advanced subject dealing with complex
physical systems and requiring sophisticated mathematics. The book
manages to present the material in an easily digestible way. Although
basic knowledge of real analysis is required, the book takes great
care to aim at a broad audience. What makes the book particularly easy
and pleasurable to read is the careful balance between the level of
technical details and the clarity and continuity in the line of
thought.

The book is written in a style that should be easily accessible to
both mathematicians and physicists.  I am convinced that it will be an
opportunity for many to enter the beautiful subject of stability of
matter and all its interesting connections to theoretical physics and
pure mathematics.

\end{document}